\documentclass{article}

\usepackage[preprint]{neurips_2025}

\usepackage[table]{xcolor} 

\definecolor{mydarkblue}{rgb}{0,0.08,0.45}
\definecolor{standardblue}{rgb}{0.149,0.580,0.871}
\definecolor{orang}{HTML}{F28C28}
\definecolor{mydarkgreen}{rgb}{0,0.6,0}
\definecolor{tabblue}{rgb}{0.122,0.467,0.706}
\definecolor{magenta}{rgb}{1,0,1}
\usepackage[colorlinks,citecolor=standardblue]{hyperref}
\hypersetup{
 colorlinks,
 citecolor=standardblue,
 linkcolor=orang,
 urlcolor=orang}
\usepackage[utf8]{inputenc} % allow utf-8 input
\usepackage[T1]{fontenc}    % use 8-bit T1 fonts
\usepackage{hyperref}       % hyperlinks
\usepackage{url}            % simple URL typesetting
\usepackage{booktabs}       % professional-quality tables
\usepackage{amsfonts}       % blackboard math symbols
\usepackage{nicefrac}       % compact symbols for 1/2, etc.
\usepackage{microtype}      % microtypography
\usepackage{amsmath}
\usepackage{graphicx}
\usepackage{wrapfig}
\usepackage{subcaption}
\usepackage{multirow}
\newcommand{\bfx}{\mathbf{x}}

\title{ProxelGen: Generating Proteins as 3D Densities}

\newcommand*\samethanks[1][\value{footnote}]{\footnotemark[#1]}
\author{
    Felix Faltings\thanks{Equal contribution.}\;\;\textsuperscript{1}
    \quad
    Hannes Stark\samethanks\;\;\textsuperscript{1}
    \quad
    Regina Barzilay\textsuperscript{1}
    \quad
    Tommi Jaakkola\textsuperscript{1}
    \\
    \textsuperscript{1}CSAIL, Massachusetts Institute of Technology\\
    \texttt{\{faltings, hstark\}@mit.edu, \{regina, tommi\}@csail.mit.edu}
}

\begin{document}

\newcommand{\felix}[1]{\textcolor{red}{Felix: #1}}
\definecolor{lightgray}{gray}{0.95}
\definecolor{darkgray}{gray}{0.9}

\maketitle

\begin{abstract}
We develop ProxelGen, a protein structure generative model that operates on 3D densities as opposed to the prevailing 3D point cloud representations. Representing proteins as voxelized densities, or \textit{proxels}, enables new tasks and conditioning capabilities. We generate proteins encoded as proxels via a 3D CNN-based VAE in conjunction with a diffusion model operating on its latent space. Compared to state-of-the-art models, ProxelGen's samples achieve higher novelty, better FID scores, and the same level of designability as the training set. ProxelGen's advantages are demonstrated in a standard motif scaffolding benchmark, and we show how 3D density-based generation allows for more flexible shape conditioning.
\end{abstract}

\section{Introduction}
\label{sec:intro}
Recent advances in protein structure generation have explored a wide array of modeling choices, ranging from architectures to generative processes \citep{yimproteinsurvey}. The focus of our work, however,  is on an equally fundamental yet unexplored aspect of structure generation: \emph{protein representation}.  Currently, the predominant representation is to express proteins as sequences of geometric features, such as frames \citep{yim2023se3diffusionmodelapplication, yim2023fastproteinbackbonegeneration, bose2024se3stochasticflowmatchingprotein}, three-dimensional coordinates of the C$\alpha$ atoms \citep{lin2024manyonedesigningscaffolding, geffner2025proteinascalingflowbasedprotein}, or all atoms \citep{Chu2023protpardelle}. This convergence on atomistic representations influences many downstream modeling choices ultimately limiting models' performance and capabilities.

\begin{figure}[th]
  \centering
  \includegraphics[width=\linewidth]{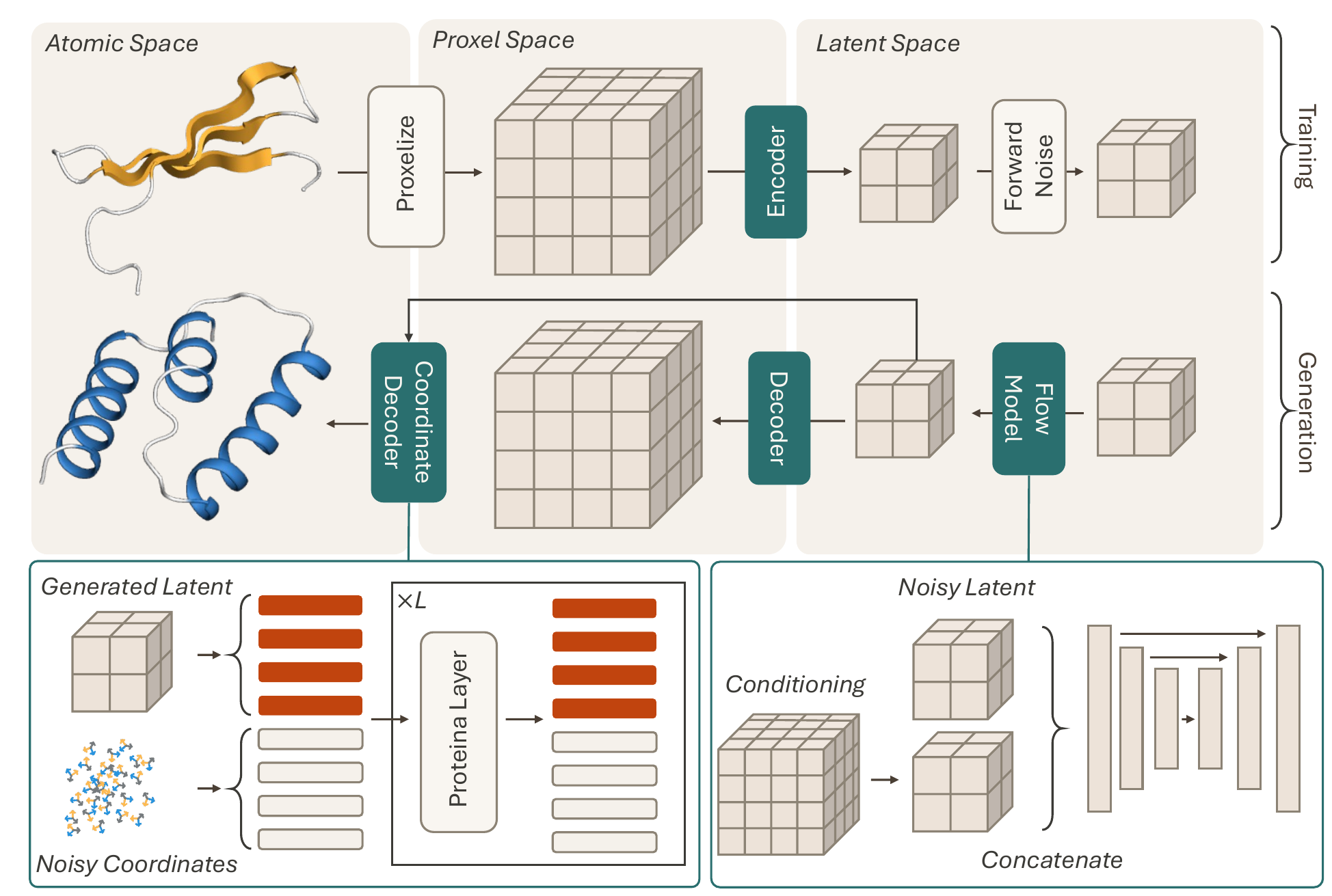}
  \caption{\textbf{Method Overview} ProxelGen trains 3 separate models. First, an \emph{autoencoder} that spatially compresses our proxel-based protein representation into latents. Second, a latent flow model trained to generate latent protein representations, which are obtained from the encoder that compresses a protein's proxel representation (encoder weights remain frozen while the flow is trained). Third, a coordinate decoder flow model that learns to decode latent voxels to atomistic representations for applications where a point cloud protein representation is desired. At inference time, the two flow models are composed to sample protein structures by first generating latent proxel representations and then decoding them via the coordinate decoder flow model.}
  \label{fig:figure1}
\end{figure}

In this paper, we explore an alternative representation of protein structure based on 3D densities called \textit{proxels} (short for \underline{pro}tein \underline{el}ements). Proxels are three-dimensional arrays, where each entry is the value of a function at a point in a three-dimensional grid, in the same way a pixel is the measured color intensity at a point in a two-dimensional grid. The proxels have multiple channels, each corresponding to a different function that encodes different information. For example, three of the channels sample Gaussian densities concentrated around the C, C$\alpha$, and N atoms in order to represent the protein backbone. To leverage this representation, we introduce ProxelGen, a latent diffusion model that directly operates on proxels. As visualized in Figure \ref{fig:figure1}, ProxelGen samples proteins encoded as proxels via a 3D CNN-based VAE in conjunction with a diffusion model operating on its latent space. 

ProxelGen takes advantage of several features of proxels that distinguish it from previous work, enabling new tasks and conditioning capabilities:
\begin{itemize}
	\item \textbf{Coarse-Graining} Proxels can be naturally coarse-grained by averaging across spatial positions. We leverage this by training ProxelGen in the spatially compressed latent space of a VAE, achieving a 512$\times$ compression in dimensions.
    \item \textbf{Convolutional Architecture} The VAE and flow model in ProxelGen use spatial convolution, imbuing them with inductive biases that are novel to the protein structure generation field. In comparison, most atomistic models use the same overarching transformer architectures operating either on frames, residues, atoms, or pairs of residues, and thus share many of the same inductive biases and disadvantages such as quadratic or cubic scaling in the length of the protein. On the other hand, ProxelGen scales with the resolution of the proxel grid, rather than the number of residues or atoms in the protein.
	\item \textbf{Spatial Conditioning} ProxelGen can naturally condition on spatial information, such as shapes or motifs specified as voxels or proxels, by directly concatenating to the model's input. For instance, proteins can be inpainted simply by masking out a region in space. For the same task, previous conditional generative models used for motif scaffolding not only require manual prespecification of the size of the protein, but also of how the designed residues map to the motif residues.
	\item \textbf{Fixed Dimensions} The dimensions of the proxels do not need to change as a function of the number of residues or chains in the protein as opposed to atomistic representations. ProxelGen thus does not require any specification of the length of the protein in advance. In contrast, the dimension of the atomistic representation is inherently tied to the size and composition of the protein, requiring the prespecification of the number of residues and chains. This issue only becomes worse in all-atom generative models, where the number of atoms is tied to the identity of the residues in the protein.
\end{itemize}

Empirically, ProxelGen's samples are more novel and show a better alignment to the training distribution than recent state-of-the-art models such as Proteina \citep{geffner2025proteinascalingflowbasedprotein}, while maintaining the same level of designability as the proteins from the AFDB training data. Moreover, we demonstrate the advantages of ProxelGen's spatial voxel-based conditioning abilities. ProxelGen is able to generate structures based on arbitrarily specified 3D shapes, and in a standard motif scaffolding task, ProxelGen improves upon baselines in tasks where the motif topology is more complicated than a simple contiguous sequence. For example, ProxelGen finds 12 unique successful designs on a 4 segment motif, as opposed to a single successful design for all other methods.

\section{Background}

\textbf{Protein Structure Generation.} 
Deep learning approaches for protein structure generation aim to aid biological discovery by proposing solutions to conditional design tasks such as motif scaffolding \citep{yim2024improvedmotifscaffoldingse3flow} and binder design \citep{Watson2023rfdiff}. 
Toward this, a crucial benchmark has been unconditional protein structure generation, based on which several axes of model design have been explored. These include architectures \citep{lin2023generatingnoveldesignablediverse, lin2024manyonedesigningscaffolding, geffner2025proteinascalingflowbasedprotein}, generative modeling paradigms and processes \citep{yim2023fastproteinbackbonegeneration, bose2024se3stochasticflowmatchingprotein}, and which geometric priors should be captured in the architecture \citep{wagner2024generatinghighlydesignableproteins, geffner2025proteinascalingflowbasedprotein} or generative process \citep{yim2023se3diffusionmodelapplication}.

Meanwhile, alternative representations for protein structures remain largely unexplored. The main extent to which variations of protein representation have been considered are variations of 3D point clouds: geometric frames \citep{yim2023se3diffusionmodelapplication}, $\alpha$-carbons \citep{lin2024manyonedesigningscaffolding, geffner2025proteinascalingflowbasedprotein}, and all-atom representations \citep{Chu2023protpardelle}.
We argue that considering a \emph{density-based} protein representation opens the door for new types of conditioning and controllable generation, favorable inductive biases, and employing optimized and scalable architectures from image generation literature.

\textbf{Density-based Generative modeling.} 
With 3D density-based modeling, we refer to representing proteins via one or multiple functions $f: \mathbb{R}^{3} \mapsto \mathbb{R}$ that assign a density to each point in 3D space. Such fields could, e.g., be the electron density or an artificially constructed density to capture the protein's shape and properties. To represent them computationally, we can discretize them into 3D grids of voxels. Operating on densities introduces a new set of possibly advantageous prior knowledge, similar to how operating on 3D point cloud representations imbues models with inductive biases that may be helpful for reasoning about biomolecules (a bias toward reasoning about distances and underlying atomic interactions). 

Employing this richer density representation as a generative modeling target has also proven fruitful in small molecule generation \citep{pinheiro2024structurebaseddrugdesigndenoising, Zhang2024ecloudgen, pinheiro20243dmoleculegenerationdenoising, dumitrescu2025e3equivariantmodelslearnchirality}. In contrast, ProxelGen explores the use of such representations for proteins for the first time. Compared to small molecules, proteins present unique challenges due to their larger size and their chain-like nature which forms an important implicit constraint on the densities.

\begin{figure}
    \centering
    \includegraphics[width=1.0\linewidth]{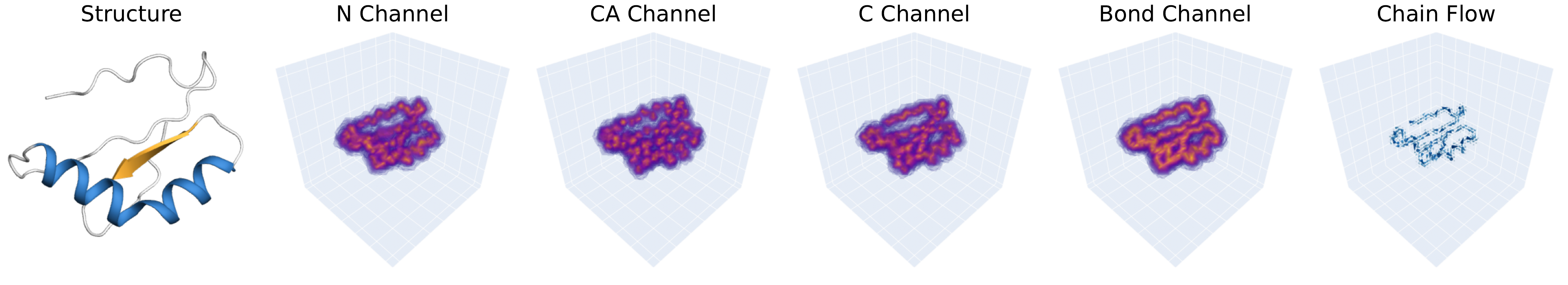}
    \caption{\textbf{Proxel Representation} A protein structure (left) is turned into a proxel representation with 7 channels: N, CA, C channels, a bond channel, and 3 channels for the chain flow.}
\end{figure}

\textbf{Latent Diffusion and Flow Models.}
Diffusion or flow models \citep{song2020score, liu2022flow, lipman2022flow, albergo2022building, albergo2023stochastic} are a class of generative models that parameterize a time-dependent vector field $v_{\theta,t}$, which can be integrated to evolve noise from a chosen prior distribution $\bfx_0 \sim p_0$ to samples from the desired data distribution $\bfx_1 \sim p_1$. Latent diffusion approaches train two models and compose them at inference time to obtain a generative model: a variational autoencoder (VAE) \citep{kingma2022autoencodingvariationalbayes} and a flow model trained to generate objects from the VAE's latent space.

Protein structure generation has struggled to obtain benefits from latent diffusion or flow approaches \citep{fu2023latentdiffusionmodelprotein, mcpartlon2024bridging, Lu2024plaid, yim2025hierarchicalproteinbackbonegeneration}, possibly due to their use of 3D atomistic representations that change their dimensionality with the number of atoms and are hard to compress into a favorable fixed-size latent space. Operating on densities discretized as 3D voxel grids instead allows ProxelGen to leverage modern, highly optimized deep learning machinery developed for 2D pixel grids (images).

\section{Method}

\newcommand{\prox}{P}

\subsection{Proxel Representation}

We represent proteins as \textit{proxel} (\underline{pro}tein \underline{el}ement) arrays: multi-channel 3D arrays $\prox\in\mathbb{R}^{C\times H\times W\times D}$, which can be seen as a discrete sampling of $C$ continuous functions on a regular grid of points in 3D. We use the different channels to represent different information about the protein. 
\paragraph{Backbone Channels} We use 3 channels to encode the positions of the C, C$\alpha$ and N backbone atoms. Taking the C$\alpha$ channels as an example, let $x_1,\dots,x_N\in\mathbb{R}^3$ be the atom coordinates. This is represented as a singular density,
\begin{equation}
    f(x) = \sum_{i=1}^N \delta_{x_i}(x).
\end{equation}
In order to expand the support of this density, it is convolved with a Gaussian kernel to give,
\begin{equation}
    g(x) = (G*f)(x) = \sum_{i=1}^NG(x-x_i),
\end{equation}
which is then sampled on a discrete grid,
\begin{equation}
    \prox^{\text{C}\alpha}_{ijk} = g(x_{ijk}),
\end{equation}
with grid points given by,
\begin{equation}
    x_{ijk} = ie_1 + je_2 + ke_3 + x_0,\quad i,j,k\in[H]\times[W]\times[D]
\end{equation}
where $x_0$ is the lower left corner of the grid and $e_1,e_2,e_3$ are the standard basis vectors.

\paragraph{Bond Channel} The fourth channel encodes the backbone bonds. For each pair of bonded backbone atoms $(x^1_i,x^2_i) \in \mathbb{R}^3\times\mathbb{R}^3$, we place density at the midpoints $\bar{x}_i = (x^1_i + x^2_i)/2$, convolve with a Gaussian, and sample on the grid as above:
\begin{equation}
	\prox^{\text{bond}}_{ijk} = g(x_{ijk}),\quad g(x) = \sum_i G(x-\bar{x}_i).
\end{equation}

\paragraph{Chain Flow} In protein design practice, the final produced protein descriptor is its sequence that can be synthesized and experimentally tested. For this purpose, we need to be able to decode our proxel representations to an \emph{ordered} sequence of amino acids. Hence, proxels and their latent features need to carry information about which density follows it in the protein's chain ordering.

We encode this as a vector field, which we term the \emph{chain flow}, that flows along the backbone from the N terminus to the C terminus. Concretely, let $v_1,\dots,v_{N-1}\in\mathbb{R}^3$ denote the sequence of vectors connecting successive C$\alpha$ positions in the backbone, $v_i = x_{i+1} - x_i$. Furthermore, let $\bar{x}_i = (x_{i+1} + x_i)/2$ denote the midpoints between C$\alpha$'s. The vector field is then,
\begin{equation}
    v(x) = \sum_{i=1}^NG(x-x_i)\cdot v_i.
\end{equation}
Because the vector field takes values in $\mathbb{R}^3$, this accounts for three channels in the voxel representation.

Finally, note that as a practical consideration, this proxelization process can be done very fast. Each density is a sum of terms that can be processed in parallel. More details are given in the appendix.

\subsection{Latent Flow Model}
\label{sec:flow_model}
\newcommand{\lat}{\ensuremath{L}}
\newcommand{\enc}{\ensuremath{E_\phi}}
\newcommand{\dec}{\ensuremath{D_\psi}}
\newcommand{\noise}{\ensuremath{Z}}
\newcommand{\model}{\ensuremath{s_\theta}}
In order to handle large proxel arrays, we follow the approach in latent diffusion \citep{rombach2022highresolutionimagesynthesislatent, vahdat2021scorebasedgenerativemodelinglatent} and learn a spatially compressed representation of the proxels. Concretely, we learn latents $\lat\in \mathbb{R}^{c\times h\times w\times d}$, where $h = H / f$, $w = W / f$, $d = D / f$ for some downsampling factor $f$ and $c\neq C$ in general. We learn $\lat$ using a $\beta$-VAE with an encoder $\enc$ which predicts the parameters of a Gaussian in latent space given proxels as input and a decoder $\dec$ which produces proxels given a latent. Letting $\mu$ denote the data distribution of proxels $\prox$, the objective is
\begin{equation}
    \mathcal{L}_{\text{AE}} = \mathbb{E}_{\lat\sim \enc(\prox),\prox\sim \mu}[\|\dec(L) - \prox\|^2_2] + \beta KL(\enc(\prox)\|\mathcal{N}(0,I)).
\end{equation}
In practice, $\beta$ is chosen to be very small in order to achieve a good reconstruction loss. Because of the sparsity of the proxels, we also found that the variances predicted by the encoder varied heavily between empty regions and regions occupied by the protein structure. This introduces an irregular geometry in the latent space, where the embedded data distribution is concentrated very tightly in some dimensions (regions occupied by the structure) and very diffuse in other dimensions (empty regions). To alleviate this, we force the distribution predicted by $\enc$ to have an identity covariance matrix. Note that this effectively reduces the KL penalty to an $L_2$ penalty on the predicted latents, preventing the latent distribution from spreading out too much, which can hinder downstream generative modeling.

After training the VAE, its weights are fixed, and we train a flow model to generate the latents $L$. We use a linear stochastic interpolant \citep{albergo2023stochastic} and train a model to predict the velocity,
\begin{equation}
	\mathcal{L}_{\text{Flow}} = \mathbb{E}_{\prox\sim \mu, \noise\sim\mathcal{N}, t\sim\nu}[\|\model(t\cdot\enc(\prox) + (1-t)\noise) - (\enc(\prox) - \noise)\|^2],
\end{equation}
where $\nu$ is a timestep distribution supported on $[0,1]$. In practice, we found it helpful to bias this distribution towards earlier times. See appendix App.~\ref{sec:app_exp_details} for details.

\subsection{Conditional Generation}

\newcommand{\cond}{\ensuremath{I}}
\newcommand{\condlat}{\ensuremath{I_L}}
We also consider conditioning on spatially structured inputs, such as masked proxels or voxelized volumes. Concretely, let $\cond\in\mathbb{R}^{H\times W\times D}$ be the conditioning input. Because our generative model operates in a spatially compressed latent space, we first map the input down to a similarly spatially compressed input $\condlat$, using a condition encoder with the same architecture as the VAE encoder $\enc$. The compressed input is then fed into the generative model by channel-wise concatenation, so that the model sees the part of the input relevant to each spatial location. The condition encoder is trained concurrently with the generative model, in contrast to the VAE encoder, which is pretrained separately.

\subsection{Structure Recovery}

Because atomic or frame representations of proteins are ubiquitous, we also consider how to decode from proxels back into a backbone representation. This allows us to, for example, use inverse folding models like ProteinMPNN to design sequences based on our generated structures and evaluate self-consistency RMSDs for comparison with previous methods.

To achieve this, we fine-tune an atomistic flow model to generate structures conditioned on proxels. Specifically, we finetune a small Proteina \citep{geffner2025proteinascalingflowbasedprotein} model to generate structures conditioned on generated proxels. To condition the model, we first tokenize the proxels by taking spatial patches, in the same way that images are tokenized in vision transformers. However, in order to reduce the number of tokens used to represent the proxels, we only tokenize the \textit{latent} representation of the proxels. We also add a 3D sinusoidal positional embedding to the proxel tokens.

These latent proxel tokens are then injected into each layer of the model by concatenating them to the regular tokens that Proteina operates on, where we also allow the proxel tokens to be updated by each layer of the model. This method avoids the need to add any additional layers into the model, which we finetune from an unconditional Proteina checkpoint. 

In order to avoid the need to finetune a much larger and more computationally demanding Proteina architecture, we opt to finetune a small 60M parameter version and then use a larger pretrained unconditional 200M parameter Proteina model to refine the decoded structures. Empirically, we found that this refinement step helps designability with only slight changes to the structures. See App.~\ref{sec:app_exp_details} for details. We hypothesize that the small model may not fully respect local geometry such as bond lengths and angles leading in turn to poorly designed sequences due to ProteinMPNN's sensitivity to backbone geometry.
 
\subsection{Self-Supervised Proxel Representations}
\label{sec:proxclr}

In order to directly evaluate the proxels generated by our model, we opt to use the Frechet Inception Distance \citep{heusel2017gans}, which compares the similarity of two distributions in a latent space (see Sec.~\ref{sec:exp_details} for more details). By comparing the samples generated by the model to the training set, we can assess how well it has learned the desired distribution. This metric has also recently been applied to evaluate protein generative models \citep{geffner2025proteinascalingflowbasedprotein,faltings2025proteinfidimprovedevaluation,lu2025assessing}. In order to adapt the FID to proxels, we learn self-supervised representations of proxels using contrastive learning. Specifically, we train a 3D ResNet using SimCLR \citep{chen2020simple} with random crops as data augmentations. We refer to this model as ProxCLR. See App.~\ref{sec:app_proxclr} for more details.

\section{Experiments}

The goals of our experiments are to (1) demonstrate that generating realistic and diverse protein structures is possible using our proxel representation and (2) demonstrate new capabilities enabled by the representation. We thus first evaluate unconditional generation, followed by two applications to motif scaffolding and shape-conditioned generation.

\subsection{Experimental Details}
\label{sec:exp_details}
We first give general experimental details on inference settings and evaluation metrics. See Appendix~\ref{sec:app_exp_details} for details on data processing, model architectures, and training. 

\paragraph{Sampling} Many protein structure generative models use some form of low-temperature sampling to optimize for designability, such as using a larger step scaling when integrating the flow field \citep{geffner2025proteinascalingflowbasedprotein}, using a lower noise scale in the sampling SDE \citep{Watson2023rfdiff}, or annealing rotations faster than translations in frame-based models \citep{bose2024se3stochasticflowmatchingprotein}. However, previous work \citep{faltings2025proteinfidimprovedevaluation} suggests that optimizing for designability may not be desirable. Moreover, better unconditional designability may not necessarily translate to better performance on relevant tasks such as motif scaffolding. Instead, we tune our model towards the FID and find that moderate step size inflation (multiplying the model predicted velocities by $1.5$) improves the FID, but overly large ones hurt the FID, likely because the model could drop parts of the distribution at low temperatures. Based on our previous observation in Sec.~\ref{sec:flow_model} that the early time steps are important for generation quality, we also integrate the vector field with a smaller time step size at early time steps so that more function evaluations are spent in the initial parts of generation.

\newcommand{\proxclr}{\ensuremath{\Phi}}
\paragraph{Evaluation Metrics} We use several metrics for evaluating generated proxels and protein structures. We describe here the less standard metrics and describe the others in App.~\ref{sec:app_exp_details}.
\begin{itemize}
	\item \textbf{FID} We evaluate the quality of generated proxels using the Frechet Inception Distance based on the self-supervised representations from our ProxCLR model $\proxclr$.  Given a sample of proxels $\prox_1,\dots,\prox_N$, and a reference set $\tilde\prox_1,\dots,\tilde\prox_M$, we compute embeddings $\proxclr(\prox_1),\dots,\proxclr(\prox_N)$ and $\proxclr(\tilde\prox_1),\dots,\proxclr(\tilde\prox_M)$. The FID is the $\mathcal{W}_2$ distance between Gaussian approximations of the two sets of embedded proxels,
	\begin{equation}
		\text{FID} = \mathcal{W}_2(\mathcal{N}(\mu_1,\Sigma_1),\mathcal{N}(\mu_2,\Sigma_2),
	\end{equation}
	where $\mu_1$, $\Sigma_1$ and $\mu_2, \Sigma_2$ are the mean and covariance matrices of $\proxclr(\prox_1),\dots,\proxclr(\prox_N)$, and $\proxclr(\tilde\prox_1),\dots,\proxclr(\tilde\prox_M)$ respectively. In our case, we use the AFDB as a reference set.
	\item \textbf{Higher Order Contacts} We also report the number of higher-order contacts in the structures as a measure of structural complexity. Two backbone positions in a structure form a contact if their C$\alpha$ atoms are within 5\AA\ of each other. The order of a contact is taken as the number of intervening secondary structures between the two residues. For example, a contact within the same secondary structure, such as between consecutive residues in an alpha helix, has an order of 0. We report the number of contacts of order greater than 0.
	\item \textbf{Secondary Structure Statistics} We also report the percentage of residues that occur in alpha helices and beta sheets.
\end{itemize}

\subsection{Unconditional Generation}
\label{sec:ucond}
We evaluate the unconditional proxel samples from ProxelGen, as well as the structures reconstructed from the generated proxels. We compare our model against native structures (Native) from AFDB and structures recovered from native proxels (Native Proxels). As a baseline, we compare against the SOTA Proteina model \citep{geffner2025proteinascalingflowbasedprotein} using both high and low temperature sampling and two model sizes. Note that Proteina was also trained on AFDB, making the FIDs comparable. For each method, we consider 256 samples. We see from the results in Table~\ref{tab:uncond} that, compared to Proteina at low temperatures, ProxelGen achieves better FID, novelty, secondary structure distributions, and number of high-order contacts, while maintaining a designability close to the training data. Sampling from Proteina at regular temperatures improves novelty and FID, but leads to worse designability compared to ProxelGen. For reference, the FID of the generated proxels themselves, before decoding back to atomic coordinates, is 6.81. Overall, we see that ProxelGen is able to generate realistic proteins.

\begin{figure}
    \centering
    \includegraphics[width=1.0\linewidth]{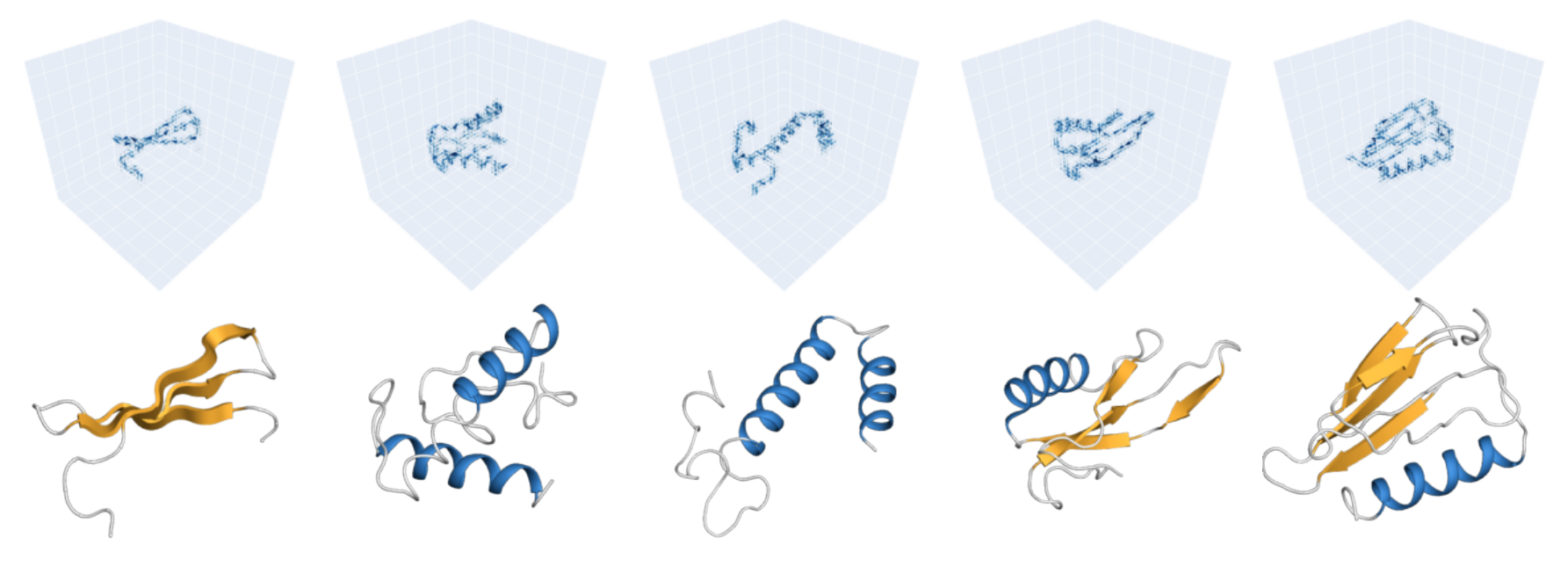}
    \caption{\textbf{Unconditional Samples} Top row: chain flow of generated proxels. Bottom row: decoded atomic structures.}
\end{figure}

\begin{table}[htb]
\centering
\caption{\textbf{Unconditional generation results} Low temperature Proteina samples marked (C) and regular temperature samples marked (H). $\alpha/\beta$ give the fraction of residues in alpha-helices/beta-sheets. Contacts gives the number of contacts of order greater than 0. Results computed on 256 samples.}
\label{tab:uncond}
\resizebox{\textwidth}{!}{%
\begin{tabular}{lllllll}
\toprule
\textbf{Sample} & \textbf{Designability} $\uparrow$ & \textbf{Diversity} $\uparrow$ & \textbf{Novelty} $\downarrow$ & \textbf{FID} $\downarrow$  & \textbf{$\alpha/\beta$} & \textbf{Contacts} \\
\midrule
\rowcolor{darkgray} Native  & 39.45  & 232  & 	0.71  & 1.89  & 34.31 / 14.99 & 29.19 \\
\rowcolor{lightgray} Native Proxels & 50.39 & 233 & 0.74 & 2.93 & 40.80 / 16.82 & 29.55 \\
\midrule 
\rowcolor{darkgray} Proteina 200M (C) & 97.77 & 127 & 0.87 & 20.47 & 69.11 / 7.08 & 15.77\\
\rowcolor{lightgray} Proteina 200M (H) & 22.27 & 243 & 0.69 & 10.53 & 33.25 / 11.90 & 15.55 \\
\rowcolor{darkgray} Proteina 400M (C) & 97.66 & 114 & 0.88 & 16.68 & 66.45 / 8.99 & 21.07 \\
\rowcolor{lightgray} Proteina 400M (H) & 45.70 & 239 & 0.74 & 7.25 & 48.00 / 12.98 & 18.77 \\
\midrule
\rowcolor{darkgray} ProxelGen & 53.13 & 233 & 0.73 & 6.05 & 41.56 / 16.05 & 24.54 \\
\bottomrule
\end{tabular}
}
\end{table}

\subsection{Inpainting and Motif Scaffolding}
\label{sec:motif_scaffolding}

\begin{table}[htb]
    \setlength{\tabcolsep}{10pt}
	\centering
	\caption{\textbf{Motif Scaffolding Results} Unique Successes on 24 motif scaffolding tasks. \textit{Free} methods do not require the specification of where the motif appears in the design, as opposed to \textit{Prespecified} ones. Lengths for free methods are averages of the successes. Lengths for prespecified methods are length ranges in the specification. Segments give the number of sequence-contiguous segments in the motif. The first four tasks contain the same motif with different length specifications. Bold values indicate row-wise maximum for tasks without different length ranges.}
    \label{tab:design25}
    \begin{tabular}{l r r r r r r}
\toprule
& & \multicolumn{1}{c}{\textbf{Free}} & \multicolumn{4}{c}{\textbf{Prespecified}}\\
\cmidrule(r){3-3} \cmidrule(lr){4-7}
\textbf{Task Name} & \rotatebox{90}{\textbf{Segments}} & \rotatebox{90}{\textbf{ProxelGen}} & \rotatebox{90}{\textbf{Proteina}} & \rotatebox{90}{\textbf{Genie2}} & \rotatebox{90}{\textbf{RFDiffusion}} & \rotatebox{90}{\textbf{FrameFlow}} \\
\midrule
\rowcolor{darkgray} 6E6R     & 1 & 34 &  &  &  & \\
\rowcolor{lightgray} \quad long & & & 713 & 415 & 381 & 110\\
\rowcolor{lightgray} \quad medium & & & 417 & 272 & 151 & 99\\
\rowcolor{lightgray} \quad short & & & 56  & 26  & 23  & 25\\
\rowcolor{darkgray} 6EXZ & 1 & 66 &  &  &  &\\
\rowcolor{lightgray} \quad long & & & 290 & 326 & 167 & 403\\
\rowcolor{lightgray} \quad medium & & & 43  & 54  & 25  & 110\\
\rowcolor{lightgray} \quad short & & & 3   & 2   & 1   & 3 \\
\rowcolor{darkgray} 5TRV     & 1 & 33 & & & & \\
\rowcolor{lightgray} \quad long  & & & 179 & 97  & 23  & 77 \\
\rowcolor{lightgray} \quad medium &  & & 22  & 23  & 10  & 21 \\
\rowcolor{lightgray} \quad short &  & & 1   & 3   & 1   & 1\\
\rowcolor{darkgray} 7MRX & 1 & 2 & & & &  \\
\rowcolor{lightgray} \quad 128 & & & 51  & 27  & 66  & 35  \\
\rowcolor{lightgray} \quad 85 & & & 31 & 23 & 13 & 22  \\
\rowcolor{lightgray} \quad 60 & & & 2   & 5   & 1   & 1\\
\midrule
\rowcolor{darkgray} 1YCR & 1 & 73 & \textbf{249} & 137 & 7  & 149  \\
\rowcolor{lightgray} 3IXT & 1 & \textbf{17} & 8   & 14  & 3   & 8  \\
\rowcolor{darkgray}5TPN & 1 & 1 & 4   & \textbf{8}   & 5   & 6  \\
\rowcolor{lightgray} 4ZYP & 1 & \textbf{22} & 11   & 3   & 6   & 4  \\
\rowcolor{darkgray} 5WN9 & 1 & 1 & 2   & 1   & 0   & \textbf{3}  \\
\rowcolor{lightgray} 1PRW & 2 & \textbf{2} & 1   & 1   & 1   & 1  \\
\rowcolor{darkgray} 5IUS & 2 & 0 & 1   & 1   & 1   & 0  \\
\rowcolor{lightgray} 2KL8 & 2 & 1 & 1   & 1   & 1   & 1 \\
\rowcolor{darkgray} 4JHW & 2 & 0 & 0   & 0   & 0   & 0  \\
\rowcolor{lightgray} 1QJG & 3 & \textbf{51} & 3   & 5   & 1   & 18  \\
\rowcolor{darkgray} 5YUI & 3 & 2 & \textbf{5}   & 3   & 1   & 1  \\
\rowcolor{lightgray} 1BCF & 4 & \textbf{12} & 1   & 1   & 1   & 1  \\
\bottomrule
\end{tabular}

    \begin{tabular}{rr}
\toprule
\multicolumn{2}{c}{\textbf{Lengths}} \\
\cmidrule{1-2}
\rotatebox{90}{\textbf{Free}} & \rotatebox{90}{\textbf{Prespecified}} \\
\midrule
\rowcolor{darkgray} 66.76 & \\
\rowcolor{lightgray}& 108 \\
\rowcolor{lightgray}& 78 \\
\rowcolor{lightgray}& 48 \\
\rowcolor{darkgray} 81.39 & \\
\rowcolor{lightgray}& 110 \\
\rowcolor{lightgray}& 80 \\
\rowcolor{lightgray}& 50 \\
\rowcolor{darkgray} 92.94 &\\
\rowcolor{lightgray}& 116 \\
\rowcolor{lightgray}& 86 \\
\rowcolor{lightgray}& 56 \\
\rowcolor{darkgray} 68.50 & \\
\rowcolor{lightgray}& 128 \\
\rowcolor{lightgray}& 85 \\
\rowcolor{lightgray}& 60 \\
\midrule
\rowcolor{darkgray} 75.82 & 40-100\\
\rowcolor{lightgray}73.88 & 50-75 \\
\rowcolor{darkgray} 52.00 & 50-75 \\
\rowcolor{lightgray} 77.36 & 30-50 \\
\rowcolor{darkgray} 23.00 & 35-50 \\
\rowcolor{lightgray} 88.00 & 60-105 \\
\rowcolor{darkgray}& 57-142 \\
\rowcolor{lightgray} 103.00 & 79 \\
\rowcolor{darkgray}& 60-90 \\
\rowcolor{lightgray} 84.80 & 53-103 \\
\rowcolor{darkgray} 116.50 & 50-100 \\
\rowcolor{lightgray} 120.33 & 96-152 \\

\bottomrule
\end{tabular}
\end{table}

We next consider how ProxelGen can be used for motif scaffolding. Similar to previous methods, we frame this as an inpainting task. However, contrary to atomistic models, the input to the model is a masked region of space, rather than a masked region of the protein sequence as in atomistic models.

We finetune a model starting from an unconditional checkpoint using the strategy from \citep{lin2024manyonedesigningscaffolding} where we randomly crop protein sequences to extract motifs that are then proxelized. When generating scaffolds, we first conditionally generate proxels and then decode them into a structure as before. When refining the structure, we take advantage of a conditional Proteina model to more strictly preserve the structure of the motif region.

We evaluate on the RFdiffusion motif scaffolding benchmark from \citep{Watson2023rfdiff}, consisting of 24 scaffolding tasks compiled from the protein design literature. For each of the 24 motifs, we design 1000 scaffolds. Each design is then scored by (1) generating 8 sequences with ProteinMPNN and (2) folding the resulting sequences with ESMFold. The designed sequences preserve the original sequence of the motif. A design is considered a success if any of the folded structures of the 8 designed sequences have (1) a backbone RMSD to the original design less than 2\AA\ (2) a motif RMSD less than 1\AA\ (3) a pLDDT greater than 70 and (4) a pAE less than 5. Successes are clustered based on a TM Score cutoff of 0.6 to give the final number of unique successes. As baselines, we compare against Proteina, Genie2 \citep{lin2024manyonedesigningscaffolding}, RFDiffusion \citep{Watson2023rfdiff}, and FrameFlow \citep{yim2023fastproteinbackbonegeneration} (results for these models reproduced from \citep{geffner2025proteinascalingflowbasedprotein}).

The results are presented in Table~\ref{tab:design25}. Because our method treats scaffolding differently, two distinctions are in order. First, \textit{free} methods denote ones that do not require a specification of where the motif should appear in the designed protein, as opposed to \textit{prespecified} methods. Second, some tasks use the same motif but with different length specifications (top half), as opposed to some tasks that only give one length specification (bottom half). For the former type of task, we only report one number for ProxelGen since we do not prespecify the lengths.

ProxelGen delivers competitive performance in motif scaffolding, achieving many more unique successes in tasks that all previous methods do poorly on, such as 1BCF or 1QJG. It is not surprising that these two motifs are composed of several disconnected segments, where the need to manually specify where each segment appears in the design is especially limiting for previous methods.

On the other hand, ProxelGen achieves fewer unique successes on other tasks, such as 6E6R. One possible explanation for the differences is the length of the designs, since longer designs are naturally more diverse, leading to more \textit{unique} successes. Indeed, the results on the first four tasks are more comparable between rows with similar lengths. However, the success of ProxelGen on 1QJG and 1BCF cannot be explained by length. In Appendix~\ref{sec:app_motif_scaffolding}, we also discuss how the lack of motif-awareness in the atomic structure decoder also limits ProxelGen's performance, particularly on larger motifs.

\begin{wrapfigure}{r}{0.4\textwidth}  % 'r' for right, 'l' for left
\vspace{-1.5cm}
  \centering
  \includegraphics[width=0.38\textwidth]{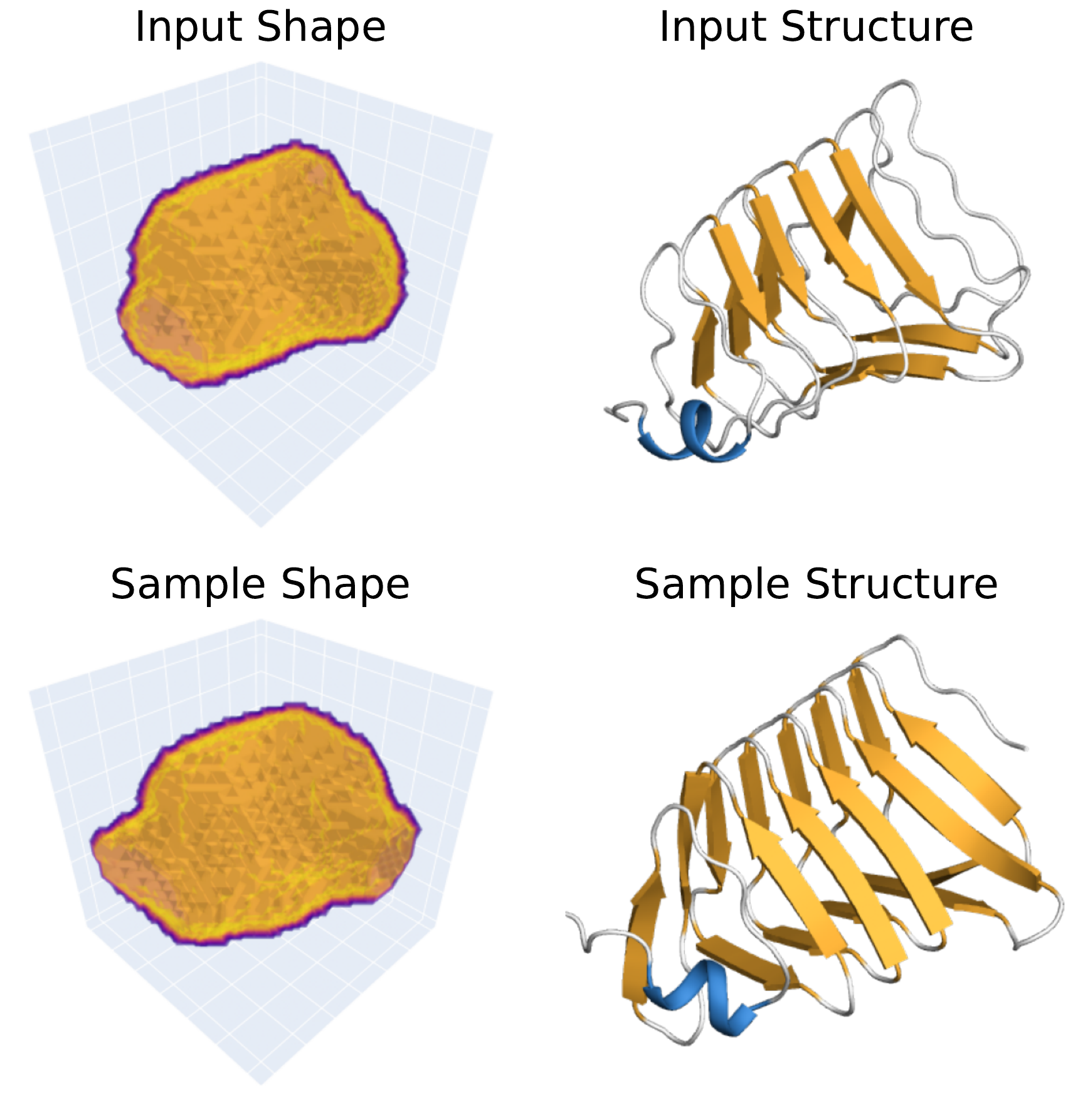}
  \caption{\textbf{Shape Conditioning} Given an input shape (top left) specified as a voxelized density derived from an input structure (top right), ProxelGen can sample a different structure (bottom right) with a similar shape. This example has a shape F1 score of $0.92$, a TM Score to the input structure of $0.36$ and a self-consistency RMSD of $0.94\AA$. }
  \vspace{-0.8cm}
\end{wrapfigure}
\subsection{Shape Conditioning}
\label{sec:shape_conditioning}

\begin{table}[tb]
	\centering
	\caption{\textbf{Shape Conditioning Results} Precision, Recall, and F1 are computed based on the input shape and the shape of the generated proteins. TMScore is computed between the generated structure and the original structure used to create the input shape.}
	\label{tab:shape_cond}
	\begin{tabular}{lrrrrr}
		\toprule
		\textbf{Subset} & \textbf{Precision} $\uparrow$ & \textbf{Recall} $\uparrow$ & \textbf{F1} $\uparrow$ & \textbf{TMScore} $\downarrow$ & \textbf{Novelty} $\downarrow$\\
		\midrule
		\rowcolor{darkgray} All & 88.55 & 88.69 & 88.60 & 60.21 & 78.28 \\
        \rowcolor{lightgray} Designable & 89.26 & 89.36 & 89.28 & 64.14 & 81.77 \\
		\bottomrule
	\end{tabular}
\end{table}

In addition to motif scaffolding, we explore other forms of spatial conditioning enabled by proxels. Here we consider conditioning on arbitrary shapes, specified as binary voxel masks. This allows us to take arbitrary geometries as input, such as protein surfaces, and produce corresponding proteins.

We create training pairs of shapes and structures by generating coarse protein surfaces. We generate the surfaces by smoothing the proxels' C$\alpha$ channel. We then sample this density on the same grid as the proxels and threshold at a single standard deviation to obtain a binary mask. As in motif scaffolding, we fine-tune a model starting from pretrained unconditional weights.

In order to evaluate the model, we consider how well it is able to generate novel proteins that conform to the same shape as existing proteins in the validation set. We evaluate a design by looking at the precision, recall, and F1 scores between the shape of the generated protein and the shape given as input. We also report the novelty and the TMScore to the original structure. As a baseline, we compared against Chroma, a structure-based model that is also able to condition on shapes.

In Table~\ref{tab:shape_cond} we report ProxelGen's results, including both all samples as well as only the designable subset. We obtained relatively poor shape adherence using Chroma (F1 $0.67$), but given the difference between Chroma's shape conditioning and ours, the results may not be fully comparable, and we therefore omitted them from the table.

We see that ProxelGen generates proteins that adhere well to the input shape while being different from the original structures. In App.~\ref{sec:app_shape_conditioning} we show that the shape adherence does not depend much on the TMScore to the original structure, indicating that ProxelGen is not simply recovering the original structures from their shapes. It is, however, possible that the model may have seen structures during training with very similar shapes despite having dissimilar structures. This is reflected in the relatively high TMScores to the training set (Novelty).

\section{Discussion}
\label{sec:discussion}
We identified an axis of protein structure generative model development along which there is a distinct lack of exploration: determining the fundamentally important question of which protein representation is the most fit for generative modeling and which benefits may arise from alternatives to the unquestionedly prevailing 3D coordinate protein representations. 
We argue for a density-based protein representation that is more true-to-nature than 3D point clouds in that it can correspond to modeling the underlying density rather than a coarse-grained atom representation. Operating on discretized voxel grids of such densities admitted using strong, established generative modeling machinery from the image generation literature, resulting in our ProxelGen model. Evaluating ProxelGen in a series of unconditional and conditional generation benchmarks showed its superior performance along several metrics while demonstrating strengths orthogonal to coordinate-based methods and enabling new types of conditioning.

For the future of density-based protein structure generation, we envision training directly on electron densities instead of employing custom-constructed densities that are derived from coordinate representations. Such electron densities are more information-rich and closer to experiment than coordinate representations, which are only derived from electron densities in an additional model-building step.  Such models, as well as the herein proposed ProxelGen, are most likely to be used for protein engineering campaigns with positive societal impacts, such as drug discovery or designing enzymes. 

\section{Limitations}
Generating proteins represented as proxels still presents many limitations. Most importantly, many of the proxels generated by our model do not form a single connected chain. For example, the chain flow may split or merge. While the coordinate decoder is still able to "thread" a protein chain through such proxels, it often introduces unnatural kinks to do so. Generating a single connected chain is also inherently challenging for convolutional models since it requires a high degree of globality. Indeed, it is not possible to check if a chain is connected by looking at local patches. This constitutes an important problem for future work.

\section{Acknowledgements}

We thank Bowen Jing, Mateo Reveiz, Rachel Wu, Sergey Ovchinnikov, Gabriele Corso, Karsten Kreis, Arash Vahdat, and Jason Yim for helpful discussions and
feedback.

We acknowledge support from the Abdul Latif Jameel Clinic for Machine Learning in Health, the NSF Expeditions grant (award 1918839: Collaborative Research: Understanding the World Through Code), the DTRA Discovery of Medical Countermeasures Against New and Emerging (DOMANE) Threats program, the MATCHMAKERS project supported by the Cancer Grand Challenges partnership 
(funded by Cancer Research UK (CGCATF-2023/100001), the National Cancer Institute (OT2CA297463) and The Mark Foundation for Cancer Research), the Machine Learning for Pharmaceutical Discovery and Synthesis (MLPDS) consortium, the Centurion Foundation and the BT Charitable Foundation.

%%%%%%%%%%%%%%%%%%%%%%%%%%%%%%%%%%%%%%%%%%%%%%%%%%%%%%
\newpage
\bibliography{references}
\bibliographystyle{plainnat}

%%%%%%%%%%%%%%%%%%%%%%%%%%%%%%%%%%%%%%%%%%%%%%%%%%%%%%%%%%%%
\newpage
\appendix

\section{Additional Technical Details}
\subsection{Proxelization} We give some additional details on how protein structures are turned into proxel representations, including how this done efficiently in practice. Recall that each channel of the proxel representation can be written as
\begin{equation}
    P^c_{ijk} = \sum_{l} w_l G(x_{ijk} - x_l),\quad i,j,k\in[H]\times[W]\times[D]
\end{equation}
for some set of $L$ points $x_l$ and weights $w_l$, and where $x_{ijk}$ is $H\times W\times D$ a grid of points. This can be parallelized in a straightfoward way by evaluating each of the $H\cdot W\cdot D \cdot L$ terms in parallel. However, for $H=W=D=32$ as in our experiments, materializing $L$ grids of $H\cdot W\cdot D=32,768$ values can become prohibitive since $L$ is usually on the order of the number of residues in a protein, which can reach 256 residues in our training data. To make this more efficient, we observe that the gaussian kernels $G(\cdot)$ we use have a standard deviation much smaller than the extent of the grid. We can thus truncate it and only evaluate each $G(x_{ijk}-x_l)$ term at points $x_{ijk}$ with some distance of $x_l$. In practice, we use a cutoff radius of $4.5$\AA\, for a Gaussian kernel with standard deviation of $1$\AA. This effectively translates to subgrids of size $5\times5\times5$ which can still be instantiated in parallel and then aggregated into the larger full-sized array. 
\subsection{Additional Experiment Details}
\label{sec:app_exp_details}
\paragraph{Timestep Training Distribution}
We empirically probed the importance of different timesteps by taking data samples, flowing them to a time $t$ and then denoising them with our generative model from that point. We found a sharp drop off in FID around $t=0.3$ indicating that the model is easily able to generate good samples provided the intermediate samples at $t=0.3$ are good. We therefore oversampled timesteps around $t=0.3$ by sampling from a mixture of a uniform distribution on $[0,1]$ and a truncated normal centered at $t=0.2$ with standard deviation $\sigma=0.1$.
\paragraph{Training Details} We use $32\times32\times32$ proxel arrays with the 7 channels described above. The discrete sampling points of the proxel grid are spaced 1.5\AA\ apart. Given that the average distance between two consecutive C$\alpha$ atoms is around 3\AA\, this ensures that two atoms are rarely placed in the same proxel cell. The Gaussian kernel we use when constructing the proxels has a standard deviation of 1\AA. 

We train on the FoldSeek \citep{Barrio-Hernandez2023fsafdb} clustered AlphaFoldDB \citep{afdb} dataset used in \citep{lin2024manyonedesigningscaffolding}, keeping only structures that fit completely on our proxel grid, leaving 98,584 structures. We randomly split this into 93,654 training structures and 4,930 validation structures.

Given the mixed advantages and disadvantages of rotation-equivariant convolutional architectures, we opt to use regular non-equivariant convolutions. We instead axis-align all the structures we train on so that their principal components align to the axes of a fixed coordinate system centered at their centers of mass. In order to avoid learning spurious biases from this procedure, we compute the alignment using perturbed structures so that, for example, nearly spherical structures are randomly aligned.

The VAE is implemented using a 3D convolutional architecture with self-attention based on the architecture used in \citep{rombach2022highresolutionimagesynthesislatent}. The VAE was trained with a KL weight of $10^{-6}$ a learning rate of $10^{-4}$ and a weight decay of $10^{-4}$. The generative model is implemented as a 3D UNet architecture, also with self-attention, based on the architecture from \citep{song2020score}. We also tried a transformer architecture but the performance was much worse. The model was trained with a learning rate of $10^{-4}$. Models were trained on single NVIDIA A6000 GPUs until relevant metrics converged (validation loss for the VAE, FID for the generative model).

\paragraph{Evaluation Metrics} Designability, diversity and novelty are standard metrics used to evaluate protein structure generative models.
\begin{itemize}
	\item \textbf{Designability} For a given protein structure, we sample 8 sequences using ProteinMPNN, which are subsequently folded with ESMFold. The structure is considered designable if the folded structure of any one of the designed sequences has an RMSD to the original structure under 2\AA.
	\item \textbf{Diversity} We compute the diversity of a set of structures by counting the number of foldseek clusters using a threshold of 0.5.
	\item \textbf{Novelty} The novelty of a structure is reported as the highest TMScore \citep{tmalign} between the structure and any structure in the training set. Note that this implies that a lower novelty is better.
\end{itemize}

\subsection{Contrastive Learning Details}
\label{sec:app_proxclr}
Let $\Phi$ denote our embedding model. Let $\prox_1,\dots,\prox_B$ denote a batch of proxelized proteins, and let $z_{2i},z_{2i-1}$ denote two augmented views of $\prox_i$. In our case, we use random cropping, rotation, and translation as augmentations. Given the cosine similarities,
\begin{equation}
	s_{i,j} = \frac{\Phi(z_i)^T\Phi(z_j)}{\|\Phi(z_i)\|\|\Phi(z_j)\|},
\end{equation}
we define the loss for pair $i,j$
\begin{equation}
	l(i,j) = -\log\frac{\exp(s_{i,j}/\tau)}{\sum_{k=1}^B\mathbb{I}_{k\neq i}\exp(s_{i,k}/\tau)}.
\end{equation}
The total loss for a batch is then,
\begin{equation}
	\mathcal{L} = \frac{1}{2B}\sum_{k=1}^B [l(2k,2k-1) + l(2k-1,2k)].
\end{equation}
We learn $\Phi$ using stochasic gradient descent following gradients of $\mathcal{L}$.

\section{Additional Results}

\subsection{Structure Refinement}
\label{sec:app_structure_refinement}
\begin{figure}
    \centering
    \includegraphics[width=0.8\linewidth]{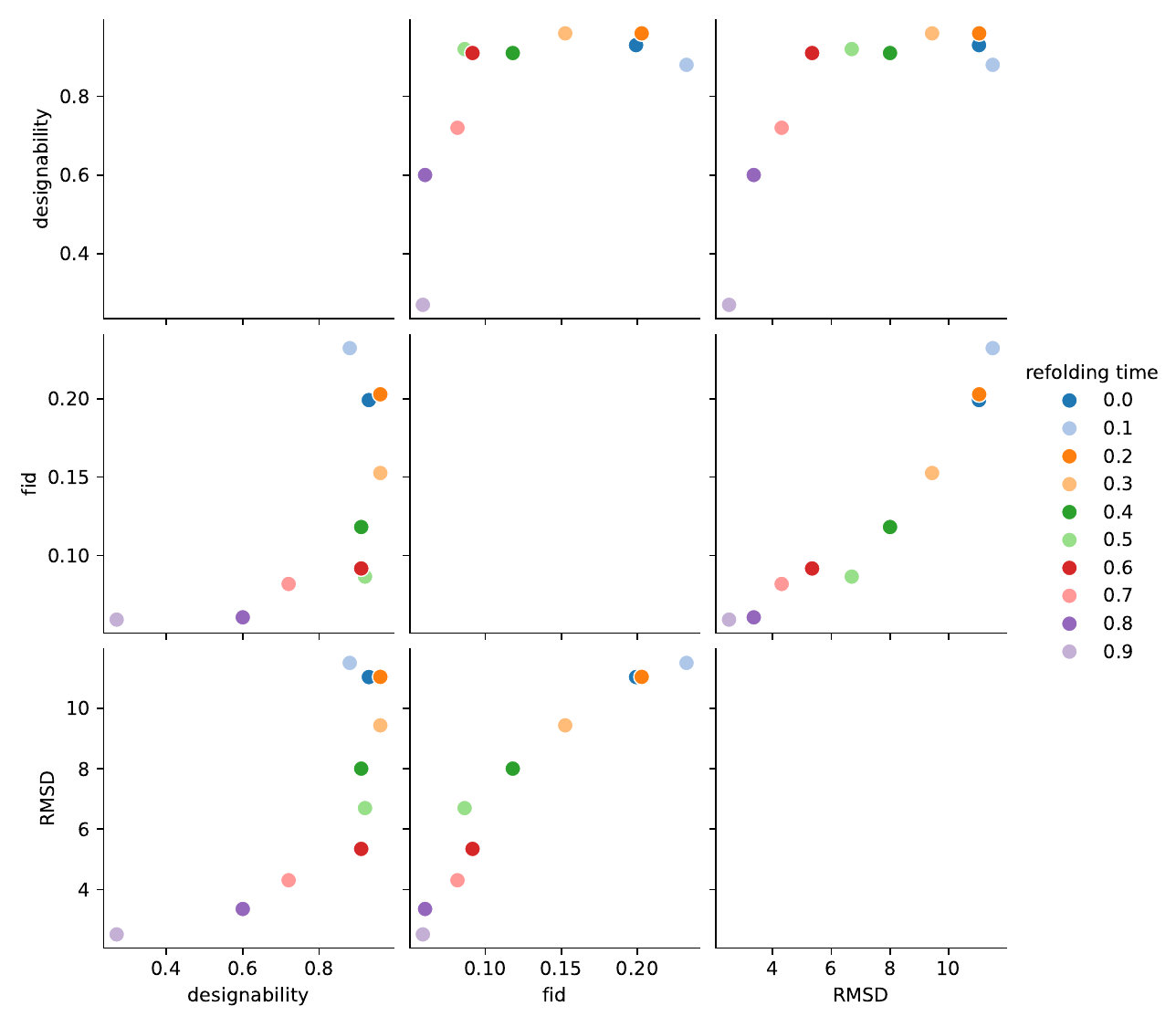}
    \caption{\textbf{Tradeoffs for Structure Refinement}: Refining the decoded atomic structures can lead to different tradeoffs between designability and fid depending on the amount of noise that is added. RMSD is computed between the refined structure and the original structure.}
    \label{fig:refinement_tradeoff}
\end{figure}

We found that the designability of the decoded structures could be vastly improved by small refinements using a much larger 200M parameter Proteina model. In order to refine structures, we renoise them down to a time $t\in[0,1]$, where $t=0$ corresponds to noise and $t=1$ to data. We then denoise using the larger model back to $t=1$. Fig.~\ref{fig:refinement_tradeoff} shows the tradeoffs between designability, FID and RMSD to the original structures for different choices of $t$. In our main experiments, we choose $t=0.8$, which gives the largest gains in designability, with little cost in FID and an average RMSD to the original structures of only around 3.35\AA.

\subsection{Motif Scaffolding}
\label{sec:app_motif_scaffolding}

Table~\ref{tab:design25_app} gives a further breakdown of ProxelGen's performance on the motif scaffolding tasks. We note that because the coordinate decoder model is not conditioned on the motif it can fail to respect the motif topology and thread the protein chain through the motif in a different way. We thus perform a first filtering of the 1000 designs from ProxelGen that checks that the motif is properly preserved. We see that this is actually a big limiter on ProxelGen's performance, particularly on larger motifs where the chance of violating the motif topology is greater. Thus performance could be greatly improved by improving the atomic decoding to be aware of the motif, though we leave this for future work.
\begin{table}[htb]
	\centering
	\caption{Additional motif scaffolding details}
    \label{tab:design25_app}
    \begin{tabular}{lrrrrrr}
\toprule
 motif & segments & mean len & motif len & unique successes & total successes & prefilter successes \\
\midrule
6e6r & 1 & 66.76 & 13 & 34 & 37 & 311 \\
6exz & 1 & 81.39 & 15 & 66 & 93 & 315 \\
5trv & 1 & 92.94 & 21 & 33 & 46 & 204 \\
7mrx & 1 & 68.50 & 22 & 2 & 5 & 94 \\
1ycr & 1 & 75.82 & 9 & 73 & 77 & 208 \\
3ixt & 1 & 73.88 & 24 & 17 & 25 & 184 \\
5tpn & 1 & 52.00 & 19 & 1 & 4 & 39 \\
4zyp & 1 & 77.36 & 15 & 22 & 26 & 132 \\
5wn9 & 1 & 23.00 & 20 & 1 & 5 & 133 \\
1prw & 2 & 88.00 & 40 & 2 & 5 & 9 \\
5ius & 2 & NaN & 42 & 0 & 0 & 0 \\
2kl8 & 2 & 103.00 & 59 & 1 & 4 & 15 \\
4jhw & 2 & NaN & 24 & 0 & 0 & 116 \\
1qjg & 3 & 84.80 & 3 & 51 & 54 & 565 \\
5yui & 3 & 116.50 & 11 & 2 & 5 & 173 \\
1bcf & 4 & 120.33 & 32 & 12 & 91 & 510 \\
\bottomrule
\end{tabular}
\end{table}

\subsection{Shape Conditioning}
\label{sec:app_shape_conditioning}

\begin{figure}
    \centering
    \includegraphics[width=1.0\linewidth]{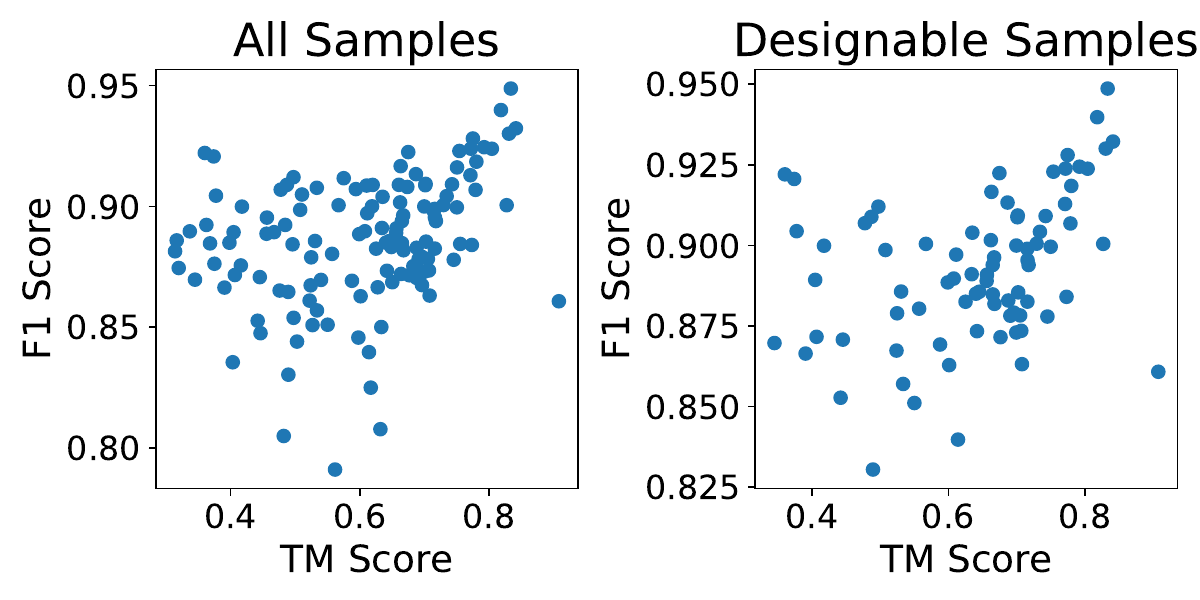}
    \caption{\textbf{Shape conditioning}: Samples with the highest TM Scores to the original structures have better shape adherence, but there are also many samples with good F1 but low TM score. Designable samples are also evenly distributed with no evident bias.}
    \label{fig:shape_cond}
\end{figure}

Figure~\ref{fig:shape_cond} plots the shape F1 score against TM Score to the original structure in our shape conditioning task. We see that while the samples with the highest TM Scores achieve slightly better shape adherence, there is no general correlation indicating that, at least on some instances, the model is not simply recovering the original structure. We also see that the designable structures are not evidently biased towards e.g. structures with higher TM Scores to the originals.

%%%%%%%%%%%%%%%%%%%%%%%%%%%%%%%%%%%%%%%%%%%%%%%%%%%%%%%%%%%%

\newpage
\section{FID Experiments}
In order to evaluate generated proxels directly without decoding back into an atomistic structure we use the FID\footnote{While the FID acronym explictly refers to the hidden representations of a specific model (Inception), we believe it is clearer to refer to all such metrics simply as FID regardless of the representations used so as to avoid creating new acronyms.}. As an embedding model, we use our self-supervised model ProxCLR (see Sec.~\ref{sec:proxclr} and App.~\ref{sec:app_proxclr}). In order to validate the use of this metric, we reproduce the experiments from \cite{faltings2025proteinfidimprovedevaluation}. These experiments consider how the FID evaluates sets of protein structures with known issues such as missing CATH clusters, or perturbed coordinates. In \cite{faltings2025proteinfidimprovedevaluation} they compute the FID using embeddings from ESM3. In our case, we first proxelize the proteins and then use representations from our ProxCLR model.

\paragraph{Effect of Perturbations} 

\begin{figure}[htbp]
  \centering

  \begin{subfigure}[b]{0.45\textwidth}
    \centering
    \includegraphics[width=\textwidth]{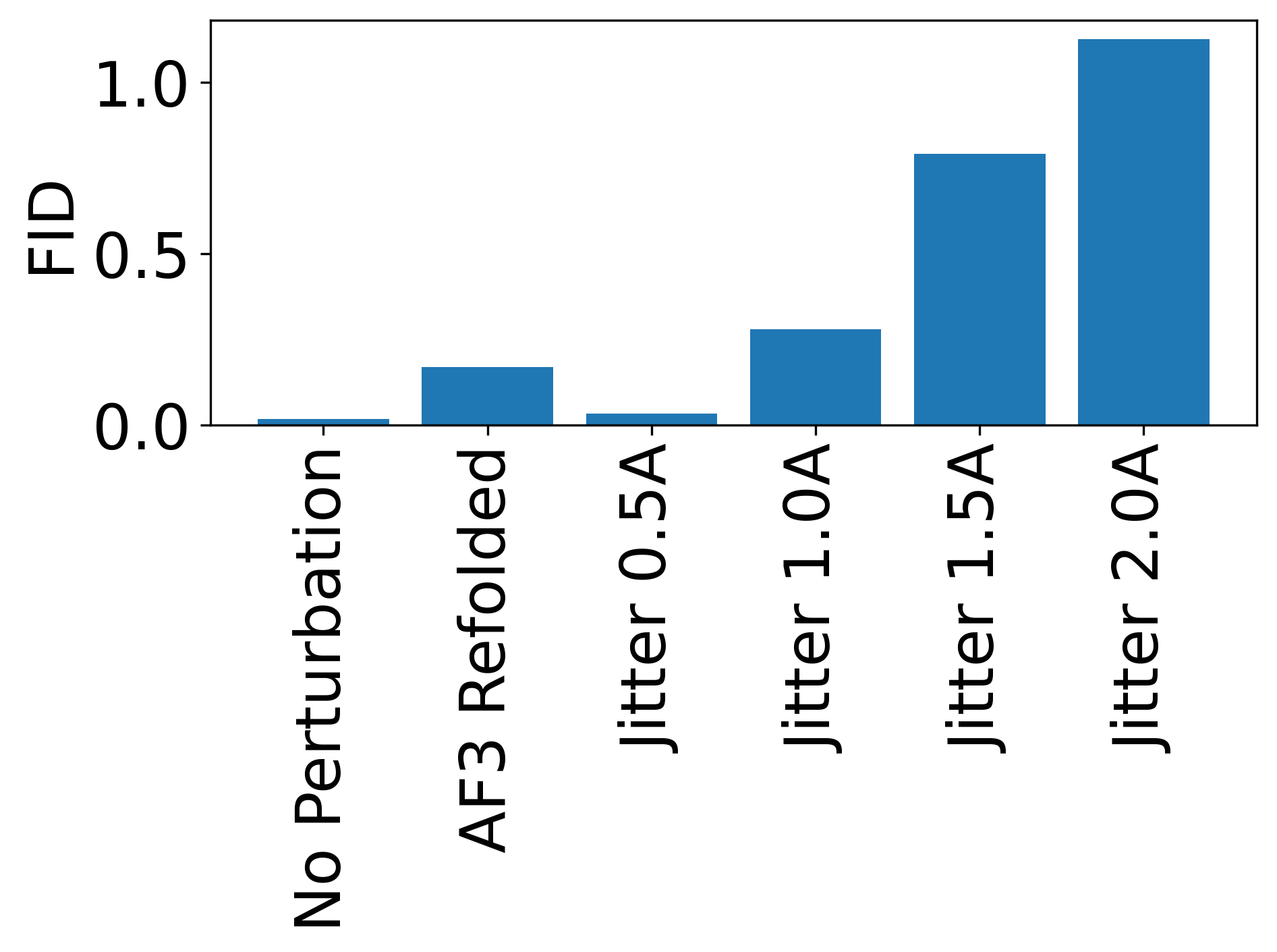}
    \caption{Effect of Perturbations}
  \end{subfigure}
  \hfill
  \begin{subfigure}[b]{0.45\textwidth}
    \centering
    \includegraphics[width=\textwidth]{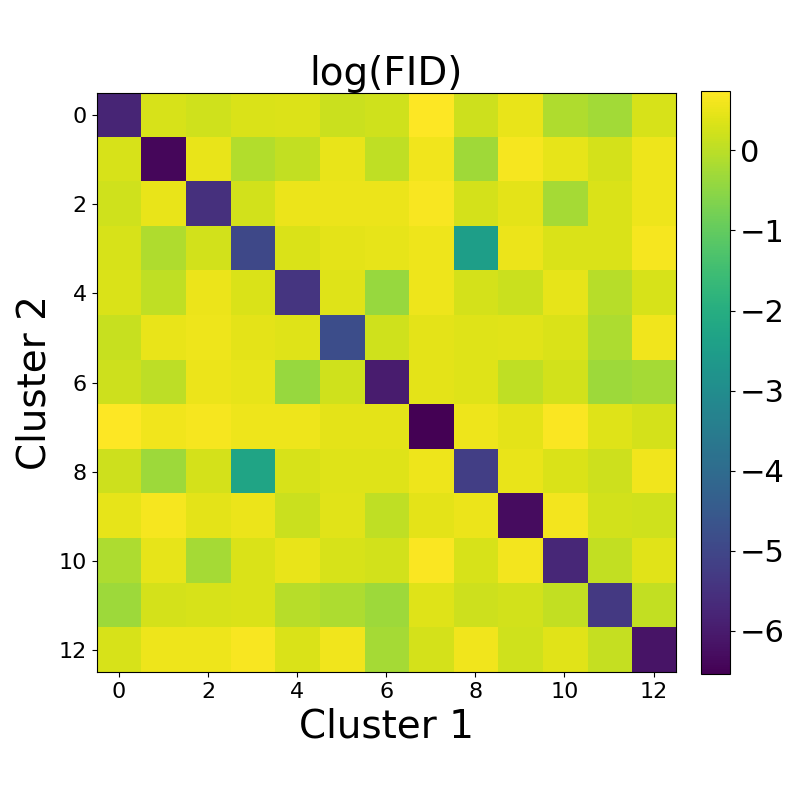}
    \caption{FID Between FoldSeek Clusters}
  \end{subfigure}

  \caption{\textbf{Perturbations and FoldSeek Cluster Comparisons} In subfigure (a) we show the FID between a reference set of PDB structures and a disjoint set of structures to which various perturbations are applied. In subfigure (b) we show the FID between disjoint samples from pairs of FoldSeek clusters. Note that on the diagonal, the two samples come from the same cluster but are disjoint.}
  \label{fig:perturbations}
\end{figure}

We first consider whether the FID is sensitive to perturbations of native structures. We use the same reference and test sets of PDB structures from \cite{faltings2025proteinfidimprovedevaluation} and the same perturbations, namely random Gaussian jittering of the atomic coordinates and refolding the structures with AlphaFold3 without MSAs. We see in Fig.~\ref{fig:perturbations} that just as with ESM3 embeddings, the FID using ProxCLR embeddings successfully detects perturbations and increases monotonically with the scale of the perturbation.

\paragraph{FoldSeek Cluster Comparisons} We also evaluate the FID between pairs of FoldSeek clusters using our ProxCLR embeddings. We see in Fig.~\ref{fig:perturbations} that the FID between disjoint samples from the same cluster is much lower than between different clusters. We also find that the FID correlates (R=$0.80$) with optimal transport distances based on TM Scores (see \cite{faltings2025proteinfidimprovedevaluation} for details).

\begin{figure}[htbp]
  \centering

  \begin{subfigure}[b]{0.45\textwidth}
    \centering
    \includegraphics[width=\textwidth]{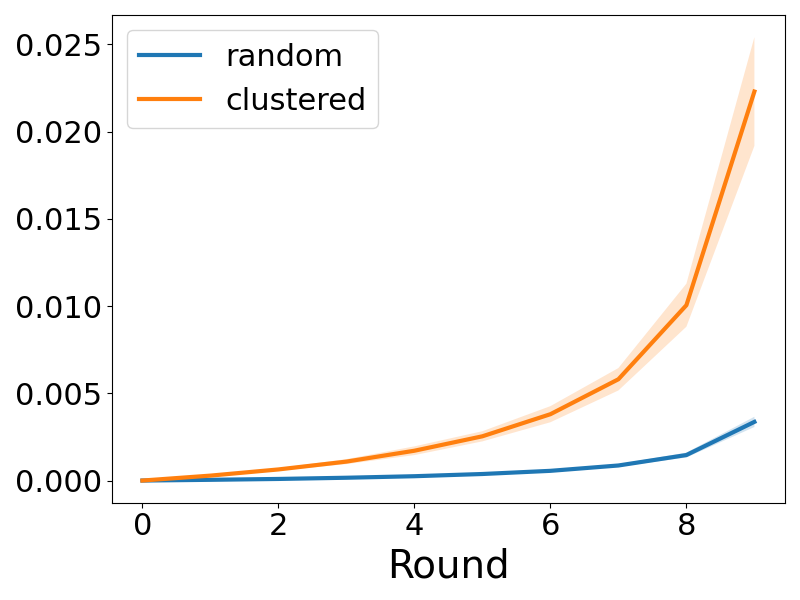}
    \caption{Effect of Removing FoldSeek Clusters}
  \end{subfigure}
  \hfill
  \begin{subfigure}[b]{0.45\textwidth}
    \centering
    \includegraphics[width=\textwidth]{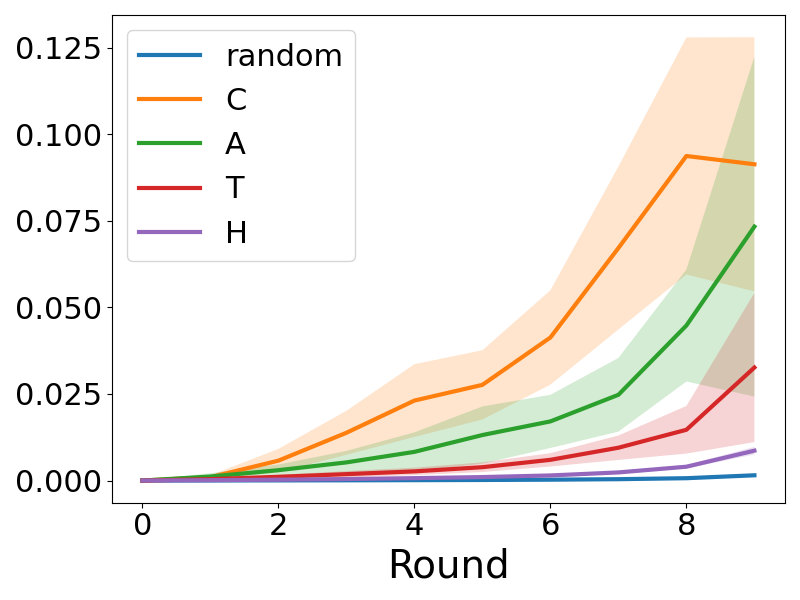}
    \caption{Effect of Removing CATH Clusters}
  \end{subfigure}

  \caption{FID sensitivity to sample diversity.}
  \label{fig:diversity_race}
\end{figure}

\paragraph{Effect of Structural Diversity} We reproduce the diversity race experiments from \cite{faltings2025proteinfidimprovedevaluation} wherein the FID is computed between a reference set and the same set from which successive subsets of structures are removed. Structures are either removed at random, or according to some clustering (either FoldSeek or CATH). We expect that removing entire clusters of structures should increase the FID more than removing structures at random. For example, removing all structures that contain $\beta$-sheets should be worse than removing the same number of structures at random, since in the former case the remaining structures will not contain any $\beta$-sheets, while in the latter case the remaining set will still be approximately representative. In Fig.~\ref{fig:diversity_race} we not only see that the FID is indeed sensitive to losing clusters of structures, but it even recovers the CATH hierarchy, as removing broader categories of structures (higher in the CATH hierarchy) hurts the FID more than removing narrower categories (lower in the hierarchy).

\end{document}